%% file: main.tex
\documentclass[twoside,leqno,twocolumn]{article}

\usepackage[letterpaper]{geometry}

\usepackage{siamproceedings}

\usepackage[T1]{fontenc}
\usepackage{amsfonts}
\usepackage{graphicx}
\usepackage{enumitem}
\usepackage{url}
\usepackage{epstopdf}
\usepackage{booktabs}
\usepackage{algpseudocode}
\usepackage{xcolor}
\usepackage{pifont}
\usepackage{caption}
\ifpdf
  \DeclareGraphicsExtensions{.eps,.pdf,.png,.jpg}
\else
  \DeclareGraphicsExtensions{.eps}
\fi

\newsiamremark{remark}{Remark}
\newsiamremark{notation}{Notation}

\begin{document}

\title{\Large Parallelizing Complex Matrix Multiplication: The 3M Solution}
\author{Yuval Spiizer\\[2pt] \normalsize Tel Aviv University\\ \normalsize \email{yuval.lahav@mail.huji.ac.il}}

\date{}

\maketitle

\fancyfoot[R]{\scriptsize{Copyright \textcopyright\ 20XX by SIAM\\
Unauthorized reproduction of this article is prohibited}}

\begin{abstract}
We consider the problem of computing the product of two matrices with complex entries, a kernel used in quantum chemistry, electromagnetic simulation, quantum circuit simulation, and FFT-based methods. Standard practice already restructures the local computation using the classical \emph{3M} (Gauss/Karatsuba) identity, computing the product from three real matrix multiplications instead of four, dropping arithmetic by $25\%$ unconditionally but leaving communication unchanged. We show that whether 3M is worth doing at all, and whether it is additionally worth restructuring communication to get more of it by computing the three products \emph{in parallel} on disjoint processor groups, are independent questions, and we answer each in closed form under a standard $\alpha$-$\beta$-$\gamma$ communication model. Under this model, for 2D layouts, in the large-$P$ limit, parallelizing the split reduces bandwidth by $13.4\%$ and latency by $1-1/\sqrt3\approx42.3\%$, the latter independent of the underlying algorithm's arithmetic exponent; neither reduction is achieved for 3D. For Strassen's algorithm the arithmetic picture reverses: restructuring locally is $7.1\%$ \emph{worse}, not better. We further generalize to Toom-Cook-$k$, giving a closed-form condition, increasing in $k$, under which higher-order splits rescue 3D layouts entirely: for instance, $k=3$ already makes 3D favorable for classical matrix multiplication, where the ordinary $k=2$ split never does. These numbers are closed-form model predictions, not measurements from a distributed implementation; we discuss what the model omits, and what would be needed to calibrate it, in \S\ref{sec:discussion}.
\end{abstract}

\begin{@abssec}{Keywords.}
Parallel Matrix Multiplication, Complex Matrix Multiplication, Communication-Avoiding Algorithms, Toom-Cook Algorithm, Strassen's Algorithm, Distributed-Memory Computing
\end{@abssec}

\input{sec-introduction}

\input{sec-preliminaries}

\input{sec-algorithms}
\input{sec-analysis}
\input{sec-strassen}

\input{sec-results}

\input{sec-toomcook}

\input{sec-bfsdfs}
\input{sec-discussion}

\section*{Acknowledgments.}
This project has received funding from the European Research Council (ERC) under the European Union's Horizon 2020 research and innovation programme (grant agreement No 818252, 101113120, 101138056). Research was supported by grant No. 1354/23 of the Israel Science Foundation (founded by the Israel Academy of Sciences and Humanities). This research has been supported by the Science Accelerator and by the Frontiers in Science initiative of the Ministry of Innovation, Science and Technology.

\bibliographystyle{siamplain}
\bibliography{references}

\end{document}

%% file: sec-introduction.tex
\section{Introduction}
\label{sec:intro}

Complex-valued matrix multiplication is a core kernel in quantum chemistry~\cite{szaboostlund,valiev2010nwchem}, electromagnetic simulation~\cite{harrington1968,taflove2005computational}, quantum circuit simulation~\cite{markovshi2008,arute2019quantum}, and any FFT-based method operating on complex data~\cite{cooleytukey1965}.

As matrix dimensions grow, computing such a product $C=A\times B$ requires distributing $A$, $B$, and $C$ across $P$ processors. This raises the general problem of \emph{parallel matrix multiplication}, studied extensively over the past four decades~\cite{cannon1969thesis,vandegeijn1997summa,choi1992scalapack,blackford1997scalapack,dekel1981,agarwal1995,mccoll1995,solomonik20112.5d,irony2004,aggarwal1990lpram,hongkung1981,valiant1990bsp,culler1993logp,bdhs2011minimizing}: how the matrices should be laid out, and how the necessary arithmetic and data movement should be scheduled, to minimize total running time on a distributed-memory machine.

Three quantities govern this running time. The \emph{arithmetic cost} is the number of scalar multiplications and additions each processor performs, as tallied for the dense matrix kernels underlying libraries such as the Level 3 BLAS~\cite{dongarra1990level3}. The \emph{communication cost}, or bandwidth, is the number of words each processor sends and receives. The \emph{latency} is the number of messages exchanged, which matters independently of message size, since every message carries a fixed overhead regardless of how much data it holds. This distinction underlies cost models such as BSP~\cite{valiant1990bsp} and LogP~\cite{culler1993logp}. A parallel algorithm's total running time is governed by all three simultaneously, and reducing one can come at the expense of another. For instance, replicating data across more processors can reduce communication at the cost of extra memory and redundant arithmetic. This paper studies how these three costs interact when a complex matrix product is computed in parallel.

Reducing communication is not a concern specific to matrix multiplication. It has been a central goal across numerical linear algebra for decades, with communication-optimal algorithms developed for QR and LU factorization~\cite{demmel2012caqr,grigori2011calu}, eigenvalue and singular value problems~\cite{ballard2011eigenvalue}, and Krylov subspace methods~\cite{hoemmen2010thesis}. Matching communication lower bounds have been established across this body of work~\cite{bdhs2011minimizing}. Toom-Cook~\cite{toom1963,cook1966} (\S\ref{sec:prelim-toomcook}) generalizes Karatsuba to $k$-way splitting. Nissim et al.\ gave Toom-Cook-$k$ algorithms that jointly optimize arithmetic and I/O~\cite{nissim2024toomcook}. Nissim and Spiizer later adapted this technique to inner-product Toom-Cook, obtaining a trade-off between arithmetic and communication costs~\cite{nissim2027innerproduct}. Bilardi and De Stefani establish a tight lower bound for the I/O complexity of Toom-Cook integer multiplication~\cite{bilardi2024toomcook}, and De Stefani analyzes the I/O complexity of sequential and parallel hybrid integer multiplication algorithms~\cite{destefani2023parallelkaratsuba}, whose parallel setting has also been separately analyzed in the fault-tolerant Toom-Cook literature~\cite{chen2024faulttoomcook,park2025faulttoomcook}.

Throughout, we compare three strategies for parallelizing the 3M identity, named once in full in \S\ref{sec:intro-contrib} below and formally defined in \S\ref{sec:algorithms}: \textbf{V1}, our proposed split into concurrent processor groups; \textbf{V2}, an in-place restructuring that keeps the standard communication schedule; and \textbf{V3}, the standard four-multiplication approach. Table~\ref{tab:numericcomparison} previews the resulting comparison.

\begin{table*}[t]
\centering
\caption{Arithmetic (V2 vs.\ V3) and communication (V1 vs.\ V2).}
\label{tab:numericcomparison}
\vspace{-6pt}
\begin{tabular*}{\textwidth}{@{\extracolsep{\fill}}llll@{}}
\toprule
Setting & V2 vs.\ V3 (arithmetic) & V1:V2 bandwidth ($P\to\infty$) & V1:V2 latency ($P\to\infty$) \\
\midrule
Classical, 2D & $-25\%$~\cite{vanzee2017complex} & $-13.4\%$ & $-42.3\%$ \\
Classical, 3D & $-25\%$~\cite{vanzee2017complex} & $+4.0\%$ & $\approx0\%$ \\
Strassen, 2D & $+7.1\%$ & $-3.7\%$ & $-42.3\%$ \\
Strassen, 3D & $+7.1\%$ & $+11.6\%$ & $\approx0\%$ \\
Karatsuba split & $-25\%$ & $-13.4\%$ & $-42.3\%$ \\
\bottomrule
\end{tabular*}
\vspace{-2pt}

{\footnotesize $k=2$, $n=2000$. Negative = reduction, positive = increase. Bandwidth and latency are the asymptotic ($P\to\infty$) V1-vs-V2 ratios of each term, from Theorem~\ref{thm:critical} and Lemma~\ref{lem:latencydominance}. Preview: this decomposition, and the two theorems behind it, are derived in \S\ref{sec:analysis}.}
\end{table*}

\subsection{Previous Work.}
\label{sec:previouswork}

The common way to compute a complex product applies the trivial formula
{\footnotesize
\begin{equation}
(a+bi)(c+di) = (ac-bd) + (ad+bc)i \label{eq:4m}
\end{equation}
}
entry-by-entry, at a cost of four real multiplications and two real additions per entry. A classical alternative needs only three multiplications: Gauss's identity for multiplying two Gaussian integers, the same identity underlying Karatsuba's algorithm~\cite{karatsuba1963}. It generalizes directly from scalars to matrix blocks $A_0,A_1,B_0,B_1$, giving the \emph{3M} identity
{\footnotesize
\begin{align}
Q_1 &= A_0B_0, \quad Q_2 = A_1B_1, \nonumber\\
Q_3 &= (A_0+A_1)(B_0+B_1), \nonumber\\
\Re(C) &= Q_1-Q_2, \label{eq:3M}\\
\Im(C) &= Q_3-Q_1-Q_2, \nonumber
\end{align}
}
at a cost of four real additions per entry instead. Entry-wise, 4M is therefore preferable, since it needs fewer total operations; 3M's advantage only appears once $Q_1,Q_2,Q_3$ are themselves large matrix products, where the multiplication it saves is a full $O(n^3)$ operation rather than a single scalar one.

This 3M/4M trade-off is well studied at the \emph{single-node} level. Van Zee et al.\ build both into BLIS, restructuring only the local, in-place computation~\cite{vanzee2017complex,vanzee2015blis}; Kouya extends this to arbitrary-precision arithmetic~\cite{arbprec2023}; the identity's mildly worse numerical stability has been separately characterized~\cite{stability3m}. A structurally identical reduction appears in the Ozaki scheme for emulating FP64 (64-bit floating-point) matrix multiplication on low-precision (FP8, 8-bit floating-point; INT8, 8-bit integer) tensor-core hardware~\cite{ozaki2012,mukunoki2020ozaki}, exposed through vendor libraries such as NVIDIA's cuBLAS~\cite{cublas}, where Uchino et al.\ recently proposed the same 4-to-3 reduction~\cite{uchino2026ozakiII}. Neither line of work considers a distributed-memory setting.

On the distributed side, communication-avoiding matrix multiplication traces back to the I/O (input/output) lower bound of Hong and Kung~\cite{hongkung1981}, generalized to distributed memory by Aggarwal et al.~\cite{aggarwal1990lpram} and Irony et al.~\cite{irony2004} and surveyed by Ballard et al. (BDHS)~\cite{bdhs2011minimizing}. The 2D, 2.5D, and 3D algorithms of \S\ref{sec:2d} realize this bound at varying memory budgets~\cite{cannon1969thesis,vandegeijn1997summa,choi1992scalapack,blackford1997scalapack,dekel1981,agarwal1995,mccoll1995,solomonik20112.5d}, alongside the BSP and LogP cost models~\cite{valiant1990bsp,culler1993logp}. In practice, 2D is used when memory is limited, since it stores only one copy of each matrix per processor; 3D is used when memory is abundant, replicating the input $P^{1/3}$-fold to cut communication substantially, and 2.5D interpolates between the two whenever partial replication fits in the available memory. Strassen's algorithm and its descendants have communication-optimal distributed implementations (CAPS, \S\ref{sec:caps})~\cite{strassen1969,winograd1971,karstadt2020little,beniamini2019,bdhs2012,lipshitz2012,ballard2013strassen}.

The distinction we draw is between the \emph{inner} real-matrix-multiplication primitive and the \emph{outer} 3M (or Toom-Cook-$k$) identity composed on top of it. The communication-avoiding and communication-optimal literature above~\cite{bdhs2012,lipshitz2012,ballard2013strassen,bdhs2011minimizing}, including CAPS, optimizes the inner primitive itself, for real- or Strassen-valued inputs; it does not address what happens when that primitive is called three (or $2k-1$) times by an outer identity, since no complex/Gaussian composition is in view. The single-node literature on 3M/4M~\cite{vanzee2017complex,uchino2026ozakiII} does consider the outer identity, but only ever restructures its \emph{local} computation (V2 in our terms), implicitly assuming the inner primitive's own communication schedule is fixed and shared; it does not consider restructuring communication around the outer identity itself, i.e., our V1. No prior work we are aware of analyzes this second, outer-level question, which is the gap this paper fills.

\subsection{Our Contributions.}
\label{sec:intro-contrib}
\label{sec:contributions}

\subsubsection{What we obtained.}
Our main contribution is a closed-form decomposition of the three-way comparison among V1, V2, and V3 (\S\ref{sec:intro}) under the standard $\alpha$-$\beta$-$\gamma$ communication model; all percentages below are predictions of that model, not measurements, and we return to what the model omits and what calibrating it would require in \S\ref{sec:discussion}. For computing the product of two complex matrices in 2D matrix-multiplication settings compared to standard practice, the model predicts an asymptotic reduction of $13.4\%$ in bandwidth and $\mathbf{42.3\%}$ in latency (Corollary~\ref{cor:trueasymptotic}). At any practically achievable scale, this extends up to a critical replication factor $c^\ast=P^{\log_3(4/3)}\approx P^{0.262}$ (Corollary~\ref{cor:criticalc}), roughly $79\%$ of the way from 2D to full 3D replication, past which in-place restructuring (V2) becomes the better choice; at astronomically large, not practically achievable $P$, splitting instead remains favorable throughout the entire 2.5D range short of full 3D replication itself (Corollary~\ref{cor:trueasymptotic-genc}).

Our second result concerns Strassen's algorithm. In 2D, splitting (V1) still reduces communication, but now at the cost of increased arithmetic: Strassen's recombination phase is taxed three times under any 3-channel split, so in-place restructuring (V2) is $\mathbf{7.1\%}$ \emph{worse} in arithmetic than standard (V3), a penalty V1 inherits. The communication saving still wins at scale: bandwidth drops by $3.7\%$ and latency by the same $42.3\%$ as classical, since the latency reduction does not depend on the arithmetic exponent, but Strassen's window is narrower (critical replication factor $P^{0.0857}$, about a third of classical's); past that, and for 3D outright, standard (V3) is the best of all three strategies at every scale we evaluated.

We generalize our technique to multi-precision matrix multiplication: applied to Karatsuba's high/low-digit split for emulating high precision on low-precision hardware, our theorems transfer unmodified, correctly predicting a classical-like (not Strassen-like) ranking. Generalizing further to $k$-digit splits (Toom-Cook-$k$), the same benefits grow with $k$ and, unlike the two-digit case, can rescue 3D layouts entirely.

Two further results round out the picture. Splitting into more pieces can rescue 3D layouts: Toom-Cook-3 suffices asymptotically for every arithmetic exponent, classical and Strassen alike, though Strassen's crossover requires a larger $P$ since its bandwidth term alone still favors standard practice at $k=3$. The same threshold generalizes to any similar splitting technique, governed by memory rather than communication for size-reducing algorithms like Strassen's own recursive step, which explains CAPS's BFS/DFS choice. We also checked whether V2's extra product forces it to use more local memory than V3: it does not. A careful choice of buffer reuse lets V2 match V3's memory footprint exactly, using only ordinary calls to an existing real-matrix-multiplication routine. We explain how we obtained all of these results next.

\subsubsection{How we obtained it.}

Once $A_0,A_1,B_0,B_1$ are distributed across $P$ processors (2D, 2.5D, or 3D, \S\ref{sec:2d}), identity~\eqref{eq:3M} admits more than one parallelization, and it is not obvious which is best. We propose parallelizing 3M by splitting processors into concurrent groups (V1 below), and compare this against the two existing alternatives (V2, V3). Three concrete strategies are possible:

\begin{enumerate}[topsep=2pt,itemsep=2pt,parsep=0pt]
\item[\textbf{V1}] \textbf{3M-split (our proposal).} Partition the $P$ processors into three groups of $P/3$; each group computes one of $Q_1,Q_2,Q_3$ concurrently.
\item[\textbf{V2}] \textbf{In-place 3M.} Keep the standard communication schedule; apply the 3M decomposition only to the local, in-place computation (as in BLIS's 3m kernel~\cite{vanzee2017complex}).
\item[\textbf{V3}] \textbf{Standard.} The common approach above: the same schedule as V2, with local computation done via the direct 4-multiplication formula.
\end{enumerate}

All five results above follow from decomposing the three-way comparison into two independent sub-questions, each answered by a single theorem. \textbf{V2 vs.\ V3} is a pure arithmetic question, since in-place restructuring (V2) and standard (V3) share, by construction, an identical communication schedule (Proposition~\ref{prop:samecomm}), so their comparison reduces to comparing two leading coefficients, $c_1^{\text{TC}}(k)=2k-1$ against $c_2(k)=k^2$ (\S\ref{sec:4m3m} and \S\ref{sec:prelim-toomcook}), independent of $n$, $P$, or the communication model. \textbf{V1 vs.\ V2} is a communication question, since splitting into $2k-1$ concurrent processor groups (V1) and in-place restructuring (V2) share, by a flop-time-invariance argument (Theorem~\ref{thm:invariance}), an identical compute time, so their comparison reduces to how bandwidth and latency scale under processor-splitting. $\theta^\ast(k)$ governs the bandwidth term alone, and hence how quickly V1 catches up to V2 as $P$ grows; the eventual winner, as $P\to\infty$, is decided by latency, which for 2D gives a universal, algorithm-independent advantage to V1 regardless of $\theta^\ast(k)$ (Corollary~\ref{cor:trueasymptotic}). Table~\ref{tab:numericcomparison} makes the two decomposed questions concrete: is 3M worth doing at all (already answered by existing practice), and is it additionally worth splitting into groups (the question this paper answers). This analysis is model-based; we also provide a consistency-check simulator as supplementary material\footnote{Anonymized code repository: \url{https://anonymous.4open.science/r/3M_mat_mul-7E4C}.}, which cross-checks our closed-form predictions against an independent, discrete accounting of each layout's communication pattern under the same cost model, catching algebraic or off-by-constant errors in the closed forms; it does not substitute for measurement on a real distributed implementation (\S\ref{sec:results}, \S\ref{sec:discussion}).

%% file: sec-preliminaries.tex
\section{Preliminaries}
\label{sec:prelim}

\subsection{Notation and Model.}
\label{sec:notation}
\label{sec:model}

An algorithm has arithmetic exponent $\omega$ if its multiplication and addition counts satisfy $M(n),A(n)=\Theta(n^\omega)$, with leading coefficients $c_M,c_A$; we call the pair $(M,A)$ its \emph{arithmetic profile}. Every algorithm here has $\omega=3$ (classical), $\omega_0=\log_27\approx2.8074$ (Strassen), or a $k$-dependent exponent (Toom-Cook-$k$). On the communication side, $BW(n,P)=n^2/P^\theta$ (bandwidth, words/processor) and $S(n,P)$ (latency, messages/processor), up to lower-order terms.

We model the machine as $P$ identical processors, connected by a peer-to-peer communication network, the standard distributed-memory model~\cite{grama2003parallel,valiant1990bsp,culler1993logp,cannon1969thesis,vandegeijn1997summa,choi1992scalapack,blackford1997scalapack,dekel1981,agarwal1995,mccoll1995,solomonik20112.5d,irony2004,aggarwal1990lpram,hongkung1981,chen2024faulttoomcook,park2025faulttoomcook}. We adopt the bilinear-algorithm formalism used throughout the communication-avoiding literature~\cite{hongkung1981,irony2004,bdhs2012,bdhs2011minimizing,nissim2024toomcook}. $C=A\times B$ is computed as $\mathrm{vec}(C)=W\big((U^\top\mathrm{vec}(A))\odot(V^\top\mathrm{vec}(B))\big)$ for encoding matrices $U,V$ and decoding matrix $W$~\cite{karstadt2020little,nissim2024toomcook}, applied to sub-blocks (Strassen) or evaluation points (Toom-Cook). Time is $T=\gamma(M+A)+\beta BW+\alpha S$ for machine constants $\gamma,\beta,\alpha$. \emph{Our aim is to reduce the leading coefficients $c_M,c_A$ and the bandwidth/latency coefficients, not merely the exponents.} These exponents are fixed once the algorithm and layout are chosen~\cite{bodrato2010,karstadt2020little,beniamini2019,nissim2024toomcook}.

Substituting a layout's memory bound into the BDHS bound~\cite{bdhs2012},
{\footnotesize
\begin{equation}
BW(n,P,M)=\Theta(n^\omega/(PM^{\omega/2-1})), \label{eq:bdhs}
\end{equation}
}
gives every bandwidth exponent $\theta(\omega)$ used throughout this paper.

\subsection{2D, 2.5D, and 3D Matrix Multiplication.}
\label{sec:2d}
\label{sec:25d}
\label{sec:3d}

All three layouts perform the same total work $n^3$, differing only in memory budget $M$ and hence communication. They span two extremes and everything in between. \textbf{2D} (Cannon~\cite{cannon1969thesis}, SUMMA~\cite{vandegeijn1997summa,choi1992scalapack,blackford1997scalapack}) is the \emph{limited-memory} extreme: $P$ processors in a $\sqrt P\times\sqrt P$ grid, minimal memory $M=\Theta(n^2/P)$, $\sqrt P$ rounds of local multiply-accumulate and shift/broadcast. \textbf{3D} (Dekel et al.~\cite{dekel1981}, refined by~\cite{agarwal1995,mccoll1995,solomonik20112.5d}) is the \emph{unlimited-memory} extreme: $P^{1/3}$-fold replication, maximal memory $M=\Theta(n^2/P^{2/3})$, local 2D multiply on each of $P^{2/3}$ layers plus a final reduction. These give $\theta_{2D}(3)=\tfrac12$, $\theta_{3D}(3)=\tfrac23$ from the BDHS bound, matching the known communication-optimal results~\cite{irony2004,solomonik20112.5d}. \textbf{2.5D}~\cite{solomonik20112.5d} interpolates via replication factor $c\in[1,P^{1/3}]$, $M=\Theta(cn^2/P)$, representing whatever memory is available between the two extremes; $c=1$ and $c=P^{1/3}$ recover 2D and 3D exactly. Substituting this $M$ into the BDHS bound~\eqref{eq:bdhs} gives $BW(n,P,c)=\Theta(n^2/\sqrt{cP})$ for classical matrix multiplication, which correctly recovers both endpoints: $n^2/\sqrt P$ at $c=1$ and $n^2/P^{2/3}$ at $c=P^{1/3}$. Table~\ref{tab:layouts} gives all three leading coefficients; replication changes \emph{only} communication, never $c_M,c_A$.

\begin{table}[t]
\centering
\caption{Communication profiles, classical matrix multiplication.}
\label{tab:layouts}
\vspace{-6pt}
\begin{tabular*}{\linewidth}{@{\extracolsep{\fill}}lll@{}}
\toprule
Layout & $BW(n,P)$ & $S(n,P)$ \\
\midrule
2D (limited $M$) & $n^2/\sqrt P$ & $\Theta(\sqrt P)$ \\
2.5D ($c$) & $\Theta(n^2/\sqrt{cP})$ & $\Theta(\sqrt{P/c^3}+\log c)$ \\
3D (unlimited $M$) & $n^2/P^{2/3}$ & $\Theta(\log P)$ \\
\bottomrule
\end{tabular*}
\vspace{-2pt}

{\footnotesize $\omega=3$; $c_M=c_A=1$ throughout, only $BW,S$ vary. Established results~\cite{cannon1969thesis,vandegeijn1997summa,dekel1981,agarwal1995,solomonik20112.5d}, re-derived here from the BDHS bound~\eqref{eq:bdhs} for consistency.}
\end{table}

\subsection{4M and 3M Gaussian Product.}
\label{sec:4m3m}

For real $A_0,A_1,B_0,B_1$, the Gaussian product $C=(A_0+iA_1)(B_0+iB_1)$ has two forms. \textbf{4M}: the common approach (Eq.~\eqref{eq:4m}), applying the trivial complex-multiplication formula entry-by-entry within a single matrix-multiplication pass. Each of the $n^3$ scalar multiply-accumulate steps forms one complex product (4 real multiplications, 2 real additions) and accumulates it into the running complex sum (2 more real additions): 4 multiplications and 4 additions per step, giving $M_{4M}(n)=A_{4M}(n)=4n^3+O(n^2)$, i.e.\ $c_M=c_A=4$. \textbf{3M} (Gauss/Karatsuba): for scalars $a,b,c,d$, the same product can instead be computed as $q_1=ac$, $q_2=bd$, $q_3=(a{+}b)(c{+}d)$, giving $\Re=q_1-q_2$ and $\Im=q_3-q_1-q_2$, using only three real multiplications instead of four. The same distributive expansion holds whether $a,b,c,d$ are scalars or matrix blocks, so this identity generalizes directly to $A_0,A_1,B_0,B_1$ (Eq.~\eqref{eq:3M}): three products plus $O(n^2)$ overhead, giving $M_{3M}(n)=A_{3M}(n)=3n^3+O(n^2)$, i.e.\ $c_M=c_A=3$.

\subsection{Toom-Cook-$k$.}
\label{sec:prelim-toomcook}

Toom-Cook-$k$~\cite{toom1963,cook1966} generalizes Karatsuba ($k=2$) to $k$-way splitting, originally for multiplying long integers: write each input as a degree-$(k{-}1)$ polynomial over $k$ digit blocks, $A'(x)=\sum_{i=0}^{k-1}A^{(i)}x^i$; the exact product is a length-$(2k{-}1)$ convolution. The schoolbook algorithm multiplies two $k$-digit integers entry-wise, at a cost of $\Theta(k^2)$ single-digit multiplications (compared to only $\Theta(k)$ operations to add two $k$-digit integers). Toom-Cook instead evaluates both inputs at $2k{-}1$ points, multiplies pointwise ($2k{-}1$ products of the reduced-size evaluations), and interpolates, using $O(k^2)$ additions for the evaluation and interpolation steps combined. Karatsuba is $k=2$ ($3$ products instead of $4$); Toom-3 is $k=3$ ($5$ instead of $9$); Toom-4 is $k=4$ ($7$ instead of $16$).

\subsection{Strassen and CAPS.}
\label{sec:prelim-strassen}
\label{sec:caps}

Strassen~\cite{strassen1969} multiplies $2\times2$ block matrices with $7$ recursive multiplications instead of $8$, at $q$ matrix additions per level (encoding + decoding), giving $M(n)=7M(n/2)$, $A(n)=7A(n/2)+\tfrac{q}{4}n^2$. Solving (verified by recursive simulation in \S\ref{sec:strassen}) gives $\omega_0=\log_27$, $c_M=1$, $c_A=q/3$. Three variants of $q$ are used throughout (Table~\ref{tab:strassen-variants}): classical, $q=18$~\cite{strassen1969}; Winograd, $q=15$~\cite{winograd1971}; Karstadt-Schwartz, $q=12$~\cite{karstadt2020little}, split as $6{+}6$ encoding/decoding. This gives $\theta_{2D}(\omega_0)\approx0.5963$, $\theta_{3D}(\omega_0)\approx0.7309$, both larger than classical's, since $\omega_0<3$ pushes $\theta$ toward the $\theta\to1$ limit at $\omega\to2$.

\textbf{CAPS}~\cite{lipshitz2012,ballard2013strassen} realizes Strassen in distributed memory by nesting its $7$-way split inside a 2.5D-style layout: as many Strassen levels as memory $M$ allows, replicating across processors, before falling back to classical 2.5D. Its arithmetic profile is exactly $c_M=1,c_A=q/3$ regardless of $P$ or replication (only communication changes, as in \S\ref{sec:25d}); it is proven communication-optimal, achieving the BDHS bound at $\omega=\omega_0$ across the same $\theta_{2D}(\omega_0)$-to-$\theta_{3D}(\omega_0)$ family as 2.5D. We use this $\theta(\omega_0)$ family directly in \S\ref{sec:strassen}, rather than re-deriving CAPS's internal engineering constants.

%% file: sec-algorithms.tex
\section{Algorithms}
\label{sec:algorithms}

We now formally define V1, V2, V3 (introduced in \S\ref{sec:intro}) and prove that each correctly computes $C=A\times B$. \S\ref{sec:analysis} analyzes their arithmetic cost, communication cost, and memory footprint.

\subsection{V1: 3M-Split (our proposal).}
Algorithm~\ref{alg:split} partitions the $P$ processors into three disjoint groups of $P/3$; each group computes one of $Q_1,Q_2,Q_3$ concurrently using the real algorithm at reduced processor count.

\begin{algorithm}[t]
\caption{V1: 3M-Split}
\label{alg:split}
\begin{algorithmic}[1]
\Require $A_0,A_1,B_0,B_1$ distributed over $P$ processors; group id $g\in\{1,2,3\}$
\State Partition the $P$ processors into 3 disjoint groups of size $P/3$
\State Scatter $(A_0,B_0)$ to group 1, $(A_1,B_1)$ to group 2, $(A_0{+}A_1,\,B_0{+}B_1)$ to group 3
\State \textbf{concurrently:} each group $g$ computes $Q_g \gets \text{RealMatMul}(\cdot,\cdot)$ using its $P/3$ processors
\State Gather $Q_1,Q_2,Q_3$; combine $\Re(C)=Q_1-Q_2,\ \Im(C)=Q_3-Q_1-Q_2$
\end{algorithmic}
\end{algorithm}

\begin{lemma}[V1 is correct]
\label{lem:v1correct}
Algorithm~\ref{alg:split} computes $C=A\times B$ exactly.
\end{lemma}
\begin{proof}
By construction, group 1 computes $Q_1=A_0B_0$, group 2 computes $Q_2=A_1B_1$, and group 3 computes $Q_3=(A_0{+}A_1)(B_0{+}B_1)$, each using only the operand blocks scattered to it. Expanding $Q_3-Q_1-Q_2=(A_0{+}A_1)(B_0{+}B_1)-A_0B_0-A_1B_1=A_0B_1+A_1B_0$ shows $\Re(C)=Q_1-Q_2=A_0B_0-A_1B_1$ and $\Im(C)=Q_3-Q_1-Q_2=A_0B_1+A_1B_0$ match the direct definition of complex multiplication exactly. Partitioning processors into disjoint groups does not affect this identity, since each $Q_g$ depends only on its own two operand blocks, which are fully available to group $g$ after the scatter step.
\end{proof}

\subsection{V2: In-place 3M.}
Algorithm~\ref{alg:v2} performs a single pass of the real algorithm's communication schedule, moving complex-valued (twice the bytes of real-valued) data, with the local, in-place computation restructured to use the 3M identity~\eqref{eq:3M} instead of the direct 4-multiplication formula: every processor locally forms $Q_1,Q_2,Q_3$ from its resident blocks of $A_0,A_1,B_0,B_1$ and combines them as $\Re(C)=Q_1-Q_2$, $\Im(C)=Q_3-Q_1-Q_2$.

\begin{algorithm}[t]
\caption{V2: In-place 3M}
\label{alg:v2}
\begin{algorithmic}[1]
\Require Local blocks $A_0,A_1,B_0,B_1$, resident identically on every one of the $P$ processors (no scatter or gather needed)
\Ensure Local blocks $\Re(C),\Im(C)$
\State $Q_1 \gets \text{RealMatMul}(A_0,B_0)$ \Comment{1st of 3 real matrix products}
\State $Q_2 \gets \text{RealMatMul}(A_1,B_1)$ \Comment{2nd of 3 real matrix products}
\State $Q_3 \gets \text{RealMatMul}(A_0{+}A_1,\ B_0{+}B_1)$ \Comment{3rd, on the two summed operands}
\State $\Re(C) \gets Q_1-Q_2$ \Comment{combine via the 3M identity~\eqref{eq:3M}}
\State $\Im(C) \gets Q_3-Q_1-Q_2$
\end{algorithmic}
\end{algorithm}

\begin{lemma}[V2 is correct]
\label{lem:v2correct}
Algorithm~\ref{alg:v2} computes $C=A\times B$ exactly.
\end{lemma}
\begin{proof}
Identical to Lemma~\ref{lem:v1correct}'s algebraic argument: V2 forms the same three products $Q_1,Q_2,Q_3$ and combines them the same way, the only difference being that all three are computed locally on each processor's already-resident data instead of being distributed across three disjoint groups. The combination step is unchanged, so correctness follows from the same expansion of $Q_3-Q_1-Q_2$.
\end{proof}

\subsection{V3: Standard.}
Algorithm~\ref{alg:v3} uses the identical communication schedule as V2 (no change to what data moves, when, or in what quantity), with local blocks combined via the direct 4-multiplication formula instead: every processor forms $\Re(C)=A_0B_0-A_1B_1$ and $\Im(C)=A_0B_1+A_1B_0$ entry-by-entry from its resident blocks, within a single pass (\S\ref{sec:intro}).

\begin{algorithm}[t]
\caption{V3: Standard}
\label{alg:v3}
\begin{algorithmic}[1]
\Require Local blocks $A_0,A_1,B_0,B_1$ (real, imaginary parts of $A,B$), resident identically on every one of the $P$ processors (same layout as V2, Algorithm~\ref{alg:v2}; no scatter or gather needed)
\Ensure Local blocks $\Re(C),\Im(C)$
\State $\Re(C),\Im(C) \gets 0$ \Comment{initialize local output blocks}
\For{each local output entry $(i,j)$}
\For{each contraction index $\ell$} \Comment{accumulate over $\ell$: the real algorithm's own schedule (\S\ref{sec:intro}), \emph{one} matrix multiplication, not 3 or 4 separate ones}
\State $a,b \gets (A_0)_{i\ell},\,(A_1)_{i\ell}$ \Comment{real, imaginary parts of $A_{i\ell}$}
\State $c,d \gets (B_0)_{\ell j},\,(B_1)_{\ell j}$ \Comment{real, imaginary parts of $B_{\ell j}$}
\State $\Re(C)_{ij} \gets \Re(C)_{ij} + ac-bd$ \Comment{trivial complex-multiply formula, 4 real mults}
\State $\Im(C)_{ij} \gets \Im(C)_{ij} + ad+bc$
\EndFor
\EndFor
\end{algorithmic}
\end{algorithm}

\begin{lemma}[V3 is correct]
\label{lem:v3correct}
Algorithm~\ref{alg:v3} computes $C=A\times B$ exactly.
\end{lemma}
\begin{proof}
Immediate from the definition of complex multiplication: expanding $(A_0+iA_1)(B_0+iB_1)$ gives $\Re(C)=A_0B_0-A_1B_1$ and $\Im(C)=A_0B_1+A_1B_0$ directly, with no intermediate identity to verify.
\end{proof}

%% file: sec-analysis.tex
\section{Analysis}
\label{sec:analysis}

We now give closed-form cost expressions for V1, V2, V3 (\S\ref{sec:algorithms}), using the arithmetic and communication leading coefficients of \S\ref{sec:prelim}. For a real-matrix-multiplication primitive of arithmetic exponent $\omega$, write $F(n,P)=n^\omega/P$ for the per-processor multiplication count (\S\ref{sec:notation}). Write $c_1$ for the leading coefficient of \emph{one} real product's combined arithmetic ($c_1=2$ for classical, by \S\ref{sec:2d}; $c_1=1+q/3$ for Strassen, by \S\ref{sec:prelim-strassen}). Write $c_2$ for the corresponding coefficient of the direct 4M (or $k^2$-product) formula applied via the standard single-pass schedule, and $c_1^{\text{TC}}$ for the 3M/Toom-Cook-$k$ analogue.

{\footnotesize
\begin{equation}
\begin{split}
T_{V1}(n,P) = {}& c_1\gamma F(n,P/3) + \beta BW(n,P/3) \\
& + \alpha S(n,P/3) + T_{\text{redist}},
\end{split} \label{eq:tv1}
\end{equation}
}
where $c_1$ here denotes the combined (mult+add) coefficient of \emph{one} real product ($c_1=2$ classical, \S\ref{sec:2d}). $T_{\text{redist}}=2\beta\frac{n^2}{P/3}+2\alpha$ approximates the one-time cost of scattering $A_0,A_1,B_0,B_1$ into their respective groups. It also covers gathering and combining $Q_1,Q_2,Q_3$ at the end, via a direct peer-to-peer exchange between corresponding processors in the three groups.

{\footnotesize
\begin{equation}
\begin{split}
T_{V2}(n,P) = {}& c_1^{\text{3M}}\,\gamma F(n,P) \\
& + 2\beta BW(n,P) + \alpha S(n,P),
\end{split} \label{eq:tv2}
\end{equation}
}
where $c_1^{\text{3M}}$ is the local-arithmetic (combined mult+add) coefficient of the 3M-decomposed computation. For classical matrix multiplication, $c_1^{\text{3M}}=3+3=6$ (\S\ref{sec:4m3m}). For Strassen, $c_1^{\text{3M}}$ is the \emph{fused-leaf} coefficient derived in \S\ref{sec:strassen}, which is \emph{not} simply $3c_1$.

{\footnotesize
\begin{equation}
\begin{split}
T_{V3}(n,P) = {}& c_2\,\gamma F(n,P) \\
& + 2\beta BW(n,P) + \alpha S(n,P).
\end{split} \label{eq:tv3}
\end{equation}
}
For classical matrix multiplication, $c_2=4$ (\S\ref{sec:4m3m}). $F(n,P)=n^3/P$ carries \emph{both} the multiplication and addition leading coefficients equally (\S\ref{sec:2d}), so the total arithmetic coefficient in~\eqref{eq:tv3} is $c_2^{M}+c_2^{A}=4+4=8$. We write this as $c_2=8$ throughout when referring to the combined (mult+add) leading coefficient, consistent with \S\ref{sec:4m3m}.

\begin{proposition}[V2 and V3 share bandwidth and latency]
\label{prop:samecomm}
$T_{V2}(n,P)-T_{V3}(n,P) = (c_1^{\text{3M}}-c_2)\gamma F(n,P)$, a pure arithmetic difference with no bandwidth or latency term.
\end{proposition}
\begin{proof}
Immediate from~\eqref{eq:tv2} and~\eqref{eq:tv3}: both use the identical $2\beta BW(n,P)+\alpha S(n,P)$ communication term, since the local restructuring from 4M to 3M changes only which arithmetic is performed on already-resident data, not what is communicated or when.
\end{proof}

Communication and arithmetic are not the only place one might expect V2 and V3 to differ: V2's identity~\eqref{eq:3M} introduces a third product, $Q_3=(A_0{+}A_1)(B_0{+}B_1)$, whose operands are not among the original four input blocks. A naive implementation that materializes both sums as brand-new temporaries before calling an existing real-matrix-multiplication routine three times needs $8$ blocks of size $n^2/P$ at peak (the $4$ inputs, $2$ outputs, and $2$ sum temporaries, all simultaneously resident at the moment $Q_3$ is computed), a $33\%$ increase over V3's $6$. This naive ordering is not forced, however.

\begin{proposition}[V2's memory footprint need not exceed V3's]
\label{prop:memory}
V2 admits an implementation with the same peak local memory footprint as V3: $6$ blocks of size $n^2/P$ (the four input blocks $A_0,A_1,B_0,B_1$ and the two output blocks $\Re(C),\Im(C)$), using only calls to an existing real-matrix-multiplication routine and no additional temporary storage.
\end{proposition}
\begin{proof}
The following order computes $C=A\times B$ using only the six buffers $A_0,A_1,B_0,B_1,\Re(C),\Im(C)$, reusing each as its original value becomes unneeded, and no others:
\begin{enumerate}
\item $\Re(C)\gets\text{RealMatMul}(A_0,B_0)$, giving $Q_1$, written directly into the output buffer.
\item $\Im(C)\gets\text{RealMatMul}(A_1,B_1)$, giving $Q_2$, written directly into the output buffer.
\item $A_0\gets A_0+A_1$ in place: $A_0$'s original value is no longer needed, since step 1 already consumed it.
\item $B_0\gets B_0+B_1$ in place: $B_0$'s original value is no longer needed, since step 1 already consumed it.
\item $A_1\gets\text{RealMatMul}(A_0,B_0)$, giving $Q_3=(A_0{+}A_1)(B_0{+}B_1)$ from the now-overwritten sum buffers, written into the now-free $A_1$ buffer; $B_0,B_1$ are not needed again.
\item $\Re(C)\gets\Re(C)-\Im(C)$, giving $Q_1-Q_2$, the correct $\Re(C)$.
\item $\Im(C)\gets A_1-\Re(C)-2\Im(C)$: since $A_1=Q_3$ and, at this point, $\Re(C)=Q_1-Q_2$ (from step 6) while $\Im(C)$ still holds $Q_2$ (unchanged by step 6), this computes $Q_3-(Q_1-Q_2)-2Q_2=Q_3-Q_1-Q_2$, the correct $\Im(C)$.
\end{enumerate}
No seventh buffer is used at any point, and each of the three real-matrix-multiplication calls (steps 1, 2, 5) is an ordinary call to an unmodified existing routine.
\end{proof}

The naive $8$-block ordering is therefore an implementation choice, not a consequence of the 3M identity itself or of reusing an existing real-matrix-multiplication routine: the reordering above achieves V3's footprint exactly, using nothing beyond ordinary buffer management in the calling code.

This decomposes the three-way comparison cleanly: V2 vs.\ V3 is a comparison of the constants $c_1^{\text{3M}}$ and $c_2$ alone (\S\ref{sec:v2v3}), while V1 vs.\ V2 is governed entirely by how communication scales under processor-splitting (\S\ref{sec:critical}). Both share the same compute time, by Theorem~\ref{thm:invariance}.

\begin{theorem}[Flop-time invariance]
\label{thm:invariance}
$T_{V1}$ and $T_{V2}$ have identical compute-time terms whenever $c_1^{\text{3M}}=3c_1$ (true for classical matrix multiplication; see \S\ref{sec:strassen} for the Strassen case, where it does not hold):
{\footnotesize
\[
3c_1\gamma F(n,P) = c_1\gamma F(n,P/3).
\]
}
\end{theorem}
\begin{proof}
$F(n,P/3)=n^\omega/(P/3)=3F(n,P)$, so $c_1\gamma F(n,P/3)=3c_1\gamma F(n,P)$.
\end{proof}
For classical matrix multiplication, Theorem~\ref{thm:invariance} isolates the V1-vs-V2 comparison to communication alone. Partitioning $P$ processors into three concurrent groups of $P/3$ costs exactly as much total processor-time as applying the 3M-restructured local computation to the entire problem at full $P$.

\subsection{V2 vs.\ V3: is 3M worth doing at all?}
\label{sec:v2v3}
For classical matrix multiplication, $c_1^{\text{3M}}=6<c_2=8$ (\S\ref{sec:4m3m}), so by Proposition~\ref{prop:samecomm}, $T_{V2}<T_{V3}$ \emph{unconditionally}. The margin is a fixed absolute quantity, $2\gamma F(n,P)$, that becomes a vanishing fraction of total time as $P\to\infty$, since the identical communication term comes to dominate, but it never reverses. This matches BLIS's real-world design choice~\cite{vanzee2017complex}. Restructuring the local kernel is free with respect to communication, so there is no reason not to do it.

For Strassen, this conclusion \emph{reverses} (\S\ref{sec:strassen}): the fused-leaf coefficient $c_1^{\text{3M}}$ exceeds $c_2$, so $T_{V2}>T_{V3}$ unconditionally; in-place 3M is strictly worse than standard, for any $P$.

\subsection{V1 vs.\ V2: is splitting worth it?}
\label{sec:critical}
\begin{theorem}[Bandwidth crossover]
\label{thm:critical}
Comparing bandwidth terms alone, V1's is smaller than V2's iff the bandwidth-coefficient ratio $3^{\theta(\omega)}$ (V1, from $BW(n,P/3)=3^{\theta(\omega)}BW(n,P)$) is less than $2$ (V2, from the complex-data doubling):
{\footnotesize
\[
3^{\theta(\omega)} < 2 \iff \theta(\omega) < \theta^\ast \equiv \log_3 2 \approx 0.6309.
\]
}
\end{theorem}
\begin{proof}
By Theorem~\ref{thm:invariance} (when it applies) the compute terms of $T_{V1}$ and $T_{V2}$ are equal, so comparing bandwidth terms in isolation, $\beta BW(n,P/3)$ against $2\beta BW(n,P)$, and solving $3^\theta=2$ for $\theta$ gives the stated threshold.
\end{proof}

This threshold governs the bandwidth terms alone, deliberately set aside from latency. It does not, by itself, determine the eventual winner: bandwidth shrinks with $P$ while latency grows, so for $P$ large enough latency dominates absolutely in both layouts, regardless of $\omega$, and the true $P\to\infty$ limit is decided by latency, not bandwidth.

\begin{lemma}[Latency dominance]
\label{lem:latencydominance}
Suppose V1 splits $P$ processors into $m$ groups (so each computes at $P/m$ processors), with a fixed redistribution overhead whose latency contribution is $2\alpha$. As $P\to\infty$, for any layout with bandwidth $BW(n,P)=\Theta(n^2/P^{\theta(\omega)})\to0$:
\begin{enumerate}
\item If $S(n,P)=\Theta(P^\rho)$ for some $\rho>0$ (polynomial growth), latency eventually dominates bandwidth in both $T_{V1}$ and $T_{V2}$, giving $T_{V1}/T_{V2}\to m^{-\rho}<1$: V1 wins, by a margin independent of $\omega$.
\item If $S(n,P)=\Theta(\log P)$ (logarithmic growth), latency again eventually dominates, but the finite absolute gap $S(n,P)-S(n,P/m)=\log_2m$ now competes against the redistribution overhead's fixed cost $2\alpha$: $T_{V1}/T_{V2}\to1$, with V1 ahead iff $\log_2m>2$.
\end{enumerate}
\end{lemma}
\begin{proof}
In case 1, $BW(n,P)\to0$ while $S(n,P)\to\infty$, so for $P$ large enough $\alpha S(n,P)$ dominates $2\beta BW(n,P)$ in $T_{V2}$, and likewise for $T_{V1}$; the ratio of latency terms is $S(n,P/m)/S(n,P)=(P/m)^\rho/P^\rho=m^{-\rho}$, giving the stated limit. In case 2, the same domination holds, but $S(n,P)-S(n,P/m)=\log_2P-\log_2(P/m)=\log_2m$ exactly, a constant; dividing $T_{V1}-T_{V2}\to(\log_2m-2)\alpha$ by $T_{V2}\to\infty$ gives $T_{V1}/T_{V2}\to1$, from below if $\log_2m>2$ and from above otherwise.
\end{proof}

\begin{corollary}[True asymptotic winner]
\label{cor:trueasymptotic}
As $P\to\infty$, $T_{V1}/T_{V2}\to\tfrac{1}{\sqrt3}\approx0.5774$ for 2D, and $\to1$ for 3D, in both cases \emph{independent of $\omega$}. V1 therefore wins asymptotically in 2D for every arithmetic exponent, by exactly $1-\tfrac{1}{\sqrt3}\approx42.3\%$; V2 remains at least as good asymptotically in 3D for every $\omega\le3$, the gap vanishing as a fraction of total time but never reversing.
\end{corollary}
\begin{proof}
Lemma~\ref{lem:latencydominance} with $m=3$: for 2D, $S(n,P)=\Theta(\sqrt P)$ is case 1 with $\rho=\tfrac12$, giving $T_{V1}/T_{V2}\to3^{-1/2}$; for 3D, $S(n,P)=\Theta(\log P)$ is case 2, and $\log_23\approx1.585<2$, so V2 remains ahead. This holds even when $\theta(\omega)>\theta^\ast$, where Theorem~\ref{thm:critical}'s bandwidth comparison alone would favor V2: bandwidth only delays the point at which V1 begins to win in 2D, since it is eventually dominated by latency regardless of its own sign.
\end{proof}

Classical 2D favors V1 at scale, converging to the universal $1-1/\sqrt3\approx42.3\%$ reduction of Corollary~\ref{cor:trueasymptotic} regardless of $\omega$; classical 3D never favors V1, converging instead to a vanishing gap in V2's favor. Theorem~\ref{thm:critical}'s bandwidth threshold still matters in both layouts, not for the sign of the eventual winner in 2D, which is unconditional, but for how large $P$ must be before that winner is realized: an algorithm with $\theta(\omega)<\theta^\ast$ (classical, Strassen) reaches the 2D asymptote at smaller $P$ than one with $\theta(\omega)>\theta^\ast$ would, though both converge to the same limit.

\begin{lemma}[V1 dominates 3M-sequential]
A strategy that runs the three products sequentially on all $P$ processors, $T_{\text{seq}}(n,P)=3(c_1\gamma F(n,P)+\beta BW(n,P)+\alpha S(n,P))$, is dominated by V1 at every $P\ge3$: $T_{V1}\le T_{\text{seq}}$, since the compute terms are equal (Theorem~\ref{thm:invariance}) while $3^{\theta(\omega)}\le3$ and the latency ratios above are both $<1$. We therefore omit 3M-sequential from the main comparison.
\end{lemma}

\subsection{The full 2.5D range: a critical replication factor.}
\label{sec:25dgeneral}
Theorem~\ref{thm:critical} compares only the two endpoints, $c=1$ and $c=P^{1/3}$. The same bandwidth-only argument extends to every intermediate $c$, giving a single number that separates "split" from "don't split" for a fixed, moderate $P$; as in \S\ref{sec:critical}, the true $P\to\infty$ limit is a separate question, taken up below.

\begin{corollary}[Critical replication factor, bandwidth-only]
\label{cor:criticalc}
Write $c=P^s$ for $s\in[0,\tfrac13]$, so $s=0$ is 2D and $s=\tfrac13$ is 3D. Substituting $M=\Theta(cn^2/P)$ into the BDHS bound~\eqref{eq:bdhs} gives a bandwidth exponent linear in $s$,
{\footnotesize
\[
\theta(s;\omega) = \left(2-\frac{\omega}{2}\right)+s\left(\frac{\omega}{2}-1\right),
\]
}
which recovers $\theta_{2D}(\omega)$ at $s=0$ and $\theta_{3D}(\omega)$ at $s=\tfrac13$. Comparing bandwidth terms alone, V1 beats V2 exactly when $c<c^\ast$, where
{\footnotesize
\[
c^\ast = P^{s^\ast(\omega)}, \qquad s^\ast(\omega) = \frac{\theta^\ast - \left(2-\frac{\omega}{2}\right)}{\frac{\omega}{2}-1}.
\]
}
\end{corollary}
\begin{proof}
Solve $\theta(s;\omega)=\theta^\ast$ for $s$, using the linear form above.
\end{proof}

For classical matrix multiplication ($\omega=3$), this simplifies to $s^\ast=\log_3(4/3)\approx0.262$, so $c^\ast=P^{\log_3(4/3)}$: splitting (V1) beats in-place restructuring (V2), comparing bandwidth alone, for any replication up to roughly $79\%$ of the way, in $\log c$, from 2D to full 3D replication.

\begin{corollary}[True asymptotic winner, general $c$]
\label{cor:trueasymptotic-genc}
As $P\to\infty$ with $s\in[0,\tfrac13)$ fixed, $T_{V1}/T_{V2}\to3^{(3s-1)/2}$, independent of $\omega$: V1 wins for \emph{every} $s<\tfrac13$, by a margin shrinking continuously to $0$ as $s\to\tfrac13$. Only at the exact endpoint $s=\tfrac13$ does this reverse, per Corollary~\ref{cor:trueasymptotic}: there V2 remains asymptotically ahead by a vanishing margin. The bandwidth-only threshold $c^\ast$ of Corollary~\ref{cor:criticalc} is therefore not the true asymptotic crossover point; there is none in the interior of $[0,\tfrac13)$, since V1 wins throughout it. $c^\ast$ instead marks, at finite $P$, where the bandwidth term's own sign changes, and remains the practically relevant threshold, since the $P$ required to observe the true asymptote grows unboundedly as $s\to\tfrac13$: at $s=0.3$, for instance, this requires $P$ beyond $10^{50}$, far past any physical system, whereas $c^\ast$ correctly predicts the finite-$P$ sign already at $P=3^{18}$.
\end{corollary}
\begin{proof}
For $s<\tfrac13$, $S(n,P,c)=\Theta(\sqrt{P/c^3})=\Theta(P^{(1-3s)/2})$: Lemma~\ref{lem:latencydominance}'s case 1 with $m=3$, $\rho=(1-3s)/2>0$, giving $T_{V1}/T_{V2}\to3^{-(1-3s)/2}=3^{(3s-1)/2}$, strictly less than $1$. As $s\to\tfrac13$, $\rho\to0$, so the $P$ at which this term overtakes bandwidth and $T_{\text{redist}}$ grows without bound, diverging exactly at $s=\tfrac13$, where $S=\Theta(\log P)$ instead and Corollary~\ref{cor:trueasymptotic}'s case 2 takes over.
\end{proof}

%% file: sec-strassen.tex
\section{Extension to Fast Matrix Multiplication}
\label{sec:strassen}

\begin{table*}[t]
\centering
\caption{Strassen arithmetic coefficients, three variants.}
\label{tab:strassen-variants}
\vspace{-6pt}
\begin{tabular*}{\textwidth}{@{\extracolsep{\fill}}lccccc@{}}
\toprule
Variant & $q$ (enc/dec) & $c_1$ (V1) & $c_2$ & $c_1^{\text{3M}}$ (V2) & V2 vs.\ V3 \\
\midrule
Classical~\cite{strassen1969} & 18 (10/8) & 7 & 18 & 19.67 & $+9.3\%$ worse \\
Winograd~\cite{winograd1971} & 15 (8/7) & 6 & 16 & 17.33 & $+8.3\%$ worse \\
Karstadt-Schwartz~\cite{karstadt2020little} & \textbf{12 (6/6)} & \textbf{5} & \textbf{14} & \textbf{15.00} & \textbf{+7.1\% worse} \\
\bottomrule
\end{tabular*}
\vspace{-2pt}

{\footnotesize $n\to\infty$. The $q$ (enc/dec) split for each variant is an established result, cited per row. V1's coefficient is $c_1$ itself, since each of its three groups computes only one real product; its comparison against V2/V3 depends on $P$ and layout, reported in Table~\ref{tab:strassen}.}
\end{table*}

Applying~\eqref{eq:bdhs} with $\omega=\omega_0=\log_27\approx2.8074$ gives $\theta_{2D}(\omega_0)\approx0.5963<\theta^\ast$ and $\theta_{3D}(\omega_0)\approx0.7309>\theta^\ast$, the same qualitative split as classical, with a \emph{larger} margin on the unfavorable 3D side ($0.10$ vs.\ classical's $0.036$).

Corollary~\ref{cor:criticalc}'s critical replication factor makes this margin concrete. Instantiating $s^\ast(\omega)$ at $\omega=\omega_0$ gives $s^\ast(\omega_0)\approx0.0857$, so $c^\ast_{\text{Strassen}}=P^{0.0857}$, compared to classical's $P^{0.262}$: Strassen's favorable window closes after only about $26\%$ of the way (in $\log c$) from 2D to full 3D replication, versus classical's $79\%$. This is the limited-memory case sharpened: on \emph{any} 2.5D layout with even modest replication, splitting for Strassen stops paying off far sooner than it does for classical matrix multiplication, since Strassen's larger $\theta_{3D}(\omega_0)$ leaves much less room in $s$ before crossing $\theta^\ast$. As Corollary~\ref{cor:trueasymptotic-genc} shows, this narrower window is a finite-$P$, bandwidth-only statement: at astronomically large $P$, splitting remains favorable for Strassen too, throughout the entire interior of the 2.5D range, the same as classical.

The arithmetic constants require more care. Recall from \S\ref{sec:prelim-strassen} that $q$ is the number of matrix additions per Strassen recursion level, split as $q=q_{\text{enc}}+q_{\text{dec}}$: encoding forms the $7$ operand pairs from $A$ and $B$'s quadrants, and decoding recombines the $7$ products into $C$'s quadrants. Solving the standard recursions gives
{\footnotesize
\[
c_1 = 1+\frac{q}{3}, \qquad c_2 = 6+\frac{2q}{3},
\]
}
for one real Strassen product and for direct-complex Strassen, respectively. The latter treats each of the $q$ additions per level as a complex operation, a factor-2 cost. Direct-complex Strassen is the natural realization of V3 for a recursive algorithm: it applies Strassen's own recursion directly to complex-typed data, with a naive complex leaf. Now consider $c_1^{\text{3M}}$, the coefficient for in-place 3M, or \emph{fused-leaf} Strassen. This carries two real channels through Strassen's recursive splitting, applies the Gaussian trick only at the scalar base case, and recombines three channels at every level. The encoding phase is paid twice (once per channel) and the decoding phase three times (once per output channel). This gives the per-level recursion $c_1^{\text{3M}}(n)=7c_1^{\text{3M}}(n/2)+\left[2q_{\text{enc}}+3q_{\text{dec}}\right](n/2)^2$, with leaf value $5$ (the Gaussian trick's $3$ multiplies $+\,2$ adds). Solving:
{\footnotesize
\[
c_1^{\text{3M}} = 5+\frac{2q_{\text{enc}}+3q_{\text{dec}}}{3}.
\]
}

Table~\ref{tab:strassen-variants} instantiates $c_1$, $c_2$, and $c_1^{\text{3M}}$ for the three Strassen variants introduced in \S\ref{sec:prelim-strassen}: classical, Winograd, and Karstadt-Schwartz. It uses the classical split $q_{\text{enc}}=10,q_{\text{dec}}=8$ and the Karstadt-Schwartz split $q_{\text{enc}}=6,q_{\text{dec}}=6$. In both cases $c_1^{\text{3M}}>c_2$: in-place 3M is worse than standard. The decoding phase's $3\times$ tax outweighs the multiplication savings, regardless of which variant's absolute addition count is used.

Using the literature-verified $q=12$ split~\cite{karstadt2020little}, $c_1=5$, $c_2=14$, $c_1^{\text{3M}}=15$: V2 (in-place 3M) is $7.1\%$ worse than V3 (standard) unconditionally, at any $P$, for any layout. This is the opposite ranking from classical matrix multiplication, and it holds regardless of which Strassen variant supplies the addition count; using the correct (rather than the commonly-assumed classical) constants roughly halves the size of the reversal but does not change its direction. CAPS (\S\ref{sec:caps}) realizes V1's split-groups strategy at Strassen's arithmetic exponent, giving communication terms $BW(n,P/3)$, $S(n,P/3)$ governed by the same $\theta_{2D}(\omega_0)$, $\theta_{3D}(\omega_0)$ derived above.

%% file: sec-results.tex
\section{Numerical Results}
\label{sec:results}

Throughout this section, all values use $\gamma=\beta=2.5\times10^{-10}$s, $\alpha=10^{-6}$s, $n=2000$. Each table reports both memory regimes of \S\ref{sec:2d}: 2D (limited memory) and 3D (unlimited memory). A negative value denotes a cost \emph{reduction} (the left-hand strategy is cheaper); a positive value denotes a cost \emph{increase}. We report V2-vs-V3 (arithmetic) and V1-vs-V2 (communication) as two separate, independent comparisons throughout, never combined into a single figure: a reduction in one is not a reduction in the other, and the two do not add. All values below were cross-checked against a consistency-check simulator provided as supplementary material: it independently derives per-processor flop counts, bytes moved, message counts, and peak memory footprint from each layout's literal communication pattern, under the same $\alpha$-$\beta$-$\gamma$ model, and checks these against the closed-form model used throughout this paper. This confirms the closed forms are algebraically consistent with the model's own assumptions; it is not a check of those assumptions against a real distributed implementation (\S\ref{sec:discussion}).

\subsection{Classical matrix multiplication.}

\begin{table}[h]
\centering
\caption{Classical matmul: arithmetic (V2:V3), communication (V1:V2 BW, latency).}
\label{tab:classical}
\vspace{-6pt}
\begin{tabular*}{\linewidth}{@{\extracolsep{\fill}}rrrrr@{}}
\toprule
Layout & $P$ & V2~\cite{vanzee2017complex}:V3 & BW & Latency \\
\midrule
2D & $8.1{\times}10^1$ & $-25.0\%$ & $-13.4\%$ & $-42.3\%$ \\
2D & $5.9{\times}10^4$ & $-13.0\%$ & $-13.4\%$ & $-42.3\%$ \\
2D & $4.8{\times}10^6$ & $-0.04\%$ & $-13.4\%$ & $-42.3\%$ \\
\midrule
3D & $8.1{\times}10^1$ & $-25.0\%$ & $+4.0\%$ & $-25.0\%$ \\
3D & $5.9{\times}10^4$ & $-23.5\%$ & $+4.0\%$ & $-10.0\%$ \\
3D & $4.8{\times}10^6$ & $-3.3\%$ & $+4.0\%$ & $-7.1\%$ \\
\bottomrule
\end{tabular*}
\vspace{-2pt}

{\footnotesize $n=2000$. From \S\ref{sec:4m3m}, Theorem~\ref{thm:critical}, Lemma~\ref{lem:latencydominance}, and Corollary~\ref{cor:trueasymptotic}.}
\end{table}

Table~\ref{tab:classical} confirms the predicted decomposition. V2-vs-V3 holds at exactly its $25\%$ reduction ($=(k-1)^2/k^2$ at $k=2$, \S\ref{sec:4m3m}) at small-to-moderate $P$ for both layouts, decaying toward $0\%$ only once communication dominates, slowly for 3D, quickly for 2D. Bandwidth is a fixed $-13.4\%$ for 2D and $+4.0\%$ for 3D at every $P$ (Theorem~\ref{thm:critical}); latency is a fixed $-42.3\%$ for 2D but shrinks toward $0\%$ for 3D as $P$ grows (Corollary~\ref{cor:trueasymptotic}). Since latency dominates as $P\to\infty$, it decides the eventual winner: split into groups on top of V2 for 2D at scale, never for 3D.

\subsection{Strassen (Karstadt-Schwartz split).}

\begin{table}[h]
\centering
\caption{Strassen algorithm: arithmetic (V2:V3), communication (V1:V2 BW, latency).}
\label{tab:strassen}
\vspace{-6pt}
\begin{tabular*}{\linewidth}{@{\extracolsep{\fill}}rrrrr@{}}
\toprule
Layout & $P$ & V2:V3 & BW & Latency \\
\midrule
2D & $8.1{\times}10^1$ & $+7.1\%$ & $-3.7\%$ & $-42.3\%$ \\
2D & $5.9{\times}10^4$ & $+2.2\%$ & $-3.7\%$ & $-42.3\%$ \\
2D & $4.3{\times}10^7$ & $+0.00\%$ & $-3.7\%$ & $-42.3\%$ \\
\midrule
3D & $8.1{\times}10^1$ & $+7.1\%$ & $+11.6\%$ & $-25.0\%$ \\
3D & $5.9{\times}10^4$ & $+6.2\%$ & $+11.6\%$ & $-10.0\%$ \\
3D & $4.3{\times}10^7$ & $+0.04\%$ & $+11.6\%$ & $-6.3\%$ \\
\bottomrule
\end{tabular*}
\vspace{-2pt}

{\footnotesize $n=2000$. From Table~\ref{tab:strassen-variants} and Corollary~\ref{cor:criticalc}.}
\end{table}

Table~\ref{tab:strassen} shows the consequence of \S\ref{sec:strassen}'s reversal: V2-vs-V3 is positive everywhere, so naively adopting in-place 3M is worse than doing nothing; a Strassen implementation built on the classical assumption needs correcting, not merely extending. Bandwidth is a fixed $-3.7\%$ for 2D and $+11.6\%$ for 3D at every $P$, larger in magnitude than classical's since Strassen's $\theta(\omega_0)$ sits further from $\theta^\ast$ in both layouts; latency, however, is exactly the same as classical's in both layouts, $\omega$-independent (Corollary~\ref{cor:trueasymptotic}): \emph{standard (V3) is the best strategy of the three for 3D-Strassen, at every scale we evaluated}. Throughout, "Strassen" refers specifically to the Karstadt-Schwartz alternative-basis technique~\cite{karstadt2020little}, the provably addition-optimal variant instantiated in Table~\ref{tab:strassen-variants}.

%% file: sec-toomcook.tex
\section{Generalizing to Multi-Digit Splits: Toom-Cook-$k$ and Precision Emulation}
\label{sec:toomcook-general}

Gauss's identity (\S\ref{sec:4m3m}) and Karatsuba's precision split (below) are both the $k=2$ case of Toom-Cook-$k$ (\S\ref{sec:prelim-toomcook}). We derive the general V2-vs-V3 and V1-vs-V2 conditions once, as a function of $k$, and instantiate both.

\subsection{General coefficients and theorems.}
Writing $c_1$ for one real product's combined (mult+add) arithmetic, $c_2(k)=k^2c_1$ and $c_1^{\text{TC}}(k)=(2k-1)c_1$ (\S\ref{sec:prelim-toomcook}).

\begin{theorem}[V2 beats V3 for every $k\ge2$]
\label{thm:generalV2V3}
For any $k\ge2$ and any base algorithm with no deep recursive recombination, $c_1^{\text{TC}}(k)<c_2(k)$, with relative saving $(k^2-(2k-1))/k^2=(k-1)^2/k^2$.
\end{theorem}
\begin{proof}
$k^2-(2k-1)=(k-1)^2>0$ for integer $k\ge2$.
\end{proof}
This generalizes the $25\%$ saving already seen at $k=2$ (\S\ref{sec:4m3m}, and Karatsuba's precision split below); the saving \emph{grows} with $k$: $44\%$ at $k=3$, $56\%$ at $k=4$, $64\%$ at $k=5$, for any algorithm without Strassen-style recursive structure (Strassen is excluded since its own $c_1$ depends on how it is split, \S\ref{sec:strassen}).

\begin{theorem}[Bandwidth crossover, general $k$]
\label{thm:generalcritical}
Comparing bandwidth terms alone, V1's is smaller than V2's iff $(2k-1)^{\theta(\omega)}<k$, i.e., $\theta(\omega)<\theta^\ast(k):=\log_{2k-1}k$.
\end{theorem}
\begin{proof}
V2 moves $k$ digit matrices (bandwidth coefficient $k$, from the complex-data-style channel doubling generalized to $k$ channels); V1 uses $P/(2k-1)$ processors per group (bandwidth coefficient $(2k-1)^{\theta(\omega)}$). Solve $(2k-1)^\theta=k$ for $\theta$. (V2 and V3 share bandwidth and latency, Proposition~\ref{prop:samecomm}, so this is equally a comparison against standard, V3.)
\end{proof}
$\theta^\ast(k)=\ln k/\ln(2k-1)$ is strictly increasing, $\theta^\ast(2)=\log_32\approx0.631$, $\theta^\ast(k)\to1$ as $k\to\infty$ (confirmed numerically: $0.631,0.683,0.712,0.733,0.747$ for $k=2,\dots,6$), since higher-order splits push the outer exponent toward the $\omega\to2$ floor. As in \S\ref{sec:critical}, this threshold governs the bandwidth terms alone: it does not by itself decide the eventual winner, since bandwidth shrinks with $P$ while latency grows, so latency eventually dominates absolutely, regardless of $\omega$.

\begin{corollary}[True asymptotic winner, general $k$]
\label{cor:trueasymptotic-genk}
As $P\to\infty$, $T_{V1}/T_{V2}\to\tfrac{1}{\sqrt{2k-1}}$ for 2D, independent of $\omega$, for every $k$ (a reduction of $1-\tfrac1{\sqrt{2k-1}}$). For 3D, $T_{V1}/T_{V2}\to1$, with the sign of the vanishing gap governed by $\log_2(2k-1)$ against $2$, also independent of $\omega$: V2 remains asymptotically ahead only at $k=2$ ($\log_23\approx1.585<2$); for every $k\ge3$ ($\log_25\approx2.322>2$), V1 is asymptotically ahead instead, however small the margin.
\end{corollary}
\begin{proof}
Lemma~\ref{lem:latencydominance} with $m=2k-1$ replacing $3$: for 2D, $\rho=\tfrac12$, giving $T_{V1}/T_{V2}\to(2k-1)^{-1/2}$; for 3D, V1 is ahead iff $\log_2(2k-1)>2\iff2k-1>4\iff k\ge3$.
\end{proof}

\begin{corollary}[Higher-order splits rescue 3D]
\label{cor:rescue3D}
For \emph{every} arithmetic exponent $\omega$, including both classical matrix multiplication and Karstadt-Schwartz Strassen, $k=2$ is asymptotically unfavorable for 3D and $k=3$ is the first to flip it favorable, by Corollary~\ref{cor:trueasymptotic-genk}. This is $\omega$-independent: it is governed by $\log_2(2k-1)$ against $2$, not by $\theta^\ast(k)$ against $\theta_{3D}(\omega)$. The bandwidth-based threshold of Theorem~\ref{thm:generalcritical} still matters for how large $P$ must be before this asymptotic winner is realized, and there it does depend on $\omega$: for classical, $\theta^\ast(3)=0.683$ already exceeds $\theta_{3D}=2/3\approx0.667$, so $k=3$'s bandwidth term alone favors V1; for Karstadt-Schwartz Strassen, $\theta^\ast(3)=0.683<\theta_{3D}(\omega_0)\approx0.731$, so $k=3$'s bandwidth term alone still favors V2, and only latency's asymptotic dominance brings V1 ahead, meaning Strassen's crossover in $k$ happens at a larger $P$ than classical's, even though both eventually cross at the same $k=3$.
\end{corollary}

\begin{table}[h]
\centering
\caption{V1:V2 bandwidth/latency, classical, 3D layout, per $k$ and $P$.}
\label{tab:toomk}
\vspace{-6pt}
\begin{tabular*}{\linewidth}{@{\extracolsep{\fill}}rccc@{}}
\toprule
$k$ & $P=10^2$ & $P=10^5$ & $P=10^8$ \\
\midrule
2 & $+4.0/{-}23.9$ & $+4.0/{-}9.5$ & $+4.0/{-}6.0$ \\
3 & $-2.5/{-}35.0$ & $-2.5/{-}14.0$ & $-2.5/{-}8.7$ \\
4 & $-8.5/{-}42.3$ & $-8.5/{-}16.9$ & $-8.5/{-}10.6$ \\
5 & $-13.5/{-}47.7$ & $-13.5/{-}19.1$ & $-13.5/{-}11.9$ \\
6 & $-17.6/{-}52.1$ & $-17.6/{-}20.8$ & $-17.6/{-}13.0$ \\
\bottomrule
\end{tabular*}
\vspace{-2pt}

{\footnotesize $n=2000$. Each cell: bandwidth$/$latency change, in \%. Negative = reduction, positive = increase. From Theorem~\ref{thm:generalcritical} and Corollary~\ref{cor:trueasymptotic-genk}.}
\end{table}

\begin{table}[h]
\centering
\caption{V1:V2 bandwidth/latency, classical, 2D layout, per $k$ and $P$.}
\label{tab:toomk2d}
\vspace{-6pt}
\begin{tabular*}{\linewidth}{@{\extracolsep{\fill}}rccc@{}}
\toprule
$k$ & $P=10^2$ & $P=10^5$ & $P=10^8$ \\
\midrule
2 & $-13.4/{-}42.3$ & $-13.4/{-}42.3$ & $-13.4/{-}42.3$ \\
3 & $-25.5/{-}55.3$ & $-25.5/{-}55.3$ & $-25.5/{-}55.3$ \\
4 & $-33.9/{-}62.2$ & $-33.9/{-}62.2$ & $-33.9/{-}62.2$ \\
5 & $-40.0/{-}66.7$ & $-40.0/{-}66.7$ & $-40.0/{-}66.7$ \\
6 & $-44.7/{-}69.8$ & $-44.7/{-}69.8$ & $-44.7/{-}69.8$ \\
\bottomrule
\end{tabular*}
\vspace{-2pt}

{\footnotesize $n=2000$. Each cell: bandwidth$/$latency change, in \%. Negative = reduction, positive = increase. From Theorem~\ref{thm:generalcritical} and Corollary~\ref{cor:trueasymptotic-genk}. Unlike 3D, both columns are constant across $P$ here, since the 2D bandwidth and latency ratios do not depend on $P$ (Theorem~\ref{thm:generalcritical}, Lemma~\ref{lem:latencydominance}).}
\end{table}

Combining both theorems into one finite-$P$ condition (V1 superior to standard iff positive):
{\footnotesize
\begin{equation}
\begin{split}
\text{save}(n,P,k) = \frac{N(n,P,k)}{k^2 + kx + y}, \quad\text{where} \\
N(n,P,k) = \big[k^2-(2k-1)\big] \\
{} + \big[k x-(2k-1)^\theta x'\big] + \big[y-y'\big],
\end{split} \label{eq:general-condition}
\end{equation}
}
with $x=\beta BW(n,P)/(\gamma c_1 F(n,P))$, $y=\alpha S(n,P)/(\gamma c_1 F(n,P))$, and $x',y'$ the same ratios at $P/(2k-1)$ processors. At finite $P$ the sign is governed jointly by the arithmetic term $(k-1)^2$ (always helps) and the communication term (helps iff $\theta(\omega)<\theta^\ast(k)$); at asymptotically large $P$, latency, not bandwidth, decides the outcome (Corollary~\ref{cor:trueasymptotic-genk}).

Table~\ref{tab:toomk} confirms Corollary~\ref{cor:rescue3D} directly, showing bandwidth and latency separately: bandwidth is negative (favorable) from $k=3$ onward and grows more favorable with $k$, while $k=2$ stays positive (unfavorable); latency is favorable throughout, for every $k$, but shrinks toward $0\%$ as $P$ grows. For classical matrix multiplication specifically, $\theta^\ast(k)$ crossing $\theta_{3D}$ already happens at $k=3$, so the bandwidth-only and true asymptotic crossovers coincide here; for algorithms where they do not (Strassen, \S\ref{sec:strassen}), the asymptotic crossover in $k$ still occurs at $k=3$, but only at a larger $P$ than the bandwidth-only threshold alone would suggest.

\subsection{The Karatsuba precision-emulation instance.}
\label{sec:karatsuba}
This framework applies unmodified to an unrelated setting: emulating a high-precision (FP64) matrix product from cheap low-precision (FP8/INT8) matrix multiplications, $100$ to $500\times$ faster per operation on recent accelerators~\cite{uchino2026ozakiII}. Writing $A'=sA^{(1)}+A^{(2)}$, $B'=sB^{(1)}+B^{(2)}$ for a base-$s$, two-digit split, Uchino et al.~\cite{uchino2026ozakiII} reduce the naive four-product formula to three, exactly the $k=2$ Toom-Cook shape with the complex $(1,i)$ basis replaced by a base-$s$ one: $A'B'=s^2C^{(1)}+C^{(2)}+s(C^{(3)}-C^{(1)}-C^{(2)})$ for $C^{(x)}{:=}A^{(x)}B^{(x)}$. Theorems~\ref{thm:generalV2V3} and~\ref{thm:generalcritical} transfer without modification: $c_1{=}2$, $c_1^{\text{3M}}{=}6$, $c_2{=}8$, and splitting is worthwhile at scale for 2D, essentially never for 3D, at the same $\theta^\ast=\log_32$.

Two caveats apply here: our evaluation uses one shared $\gamma$, though the identity's actual purpose is accessing a smaller low-precision $\gamma$, so the true crossover $P^\ast$ likely sits smaller than our numbers suggest; and the recombination carries the same cancellation risk Higham~\cite{stability3m} documents for the Gaussian case, not yet checked independently for this instance, though a general error analysis of the broader Ozaki-II scheme has recently appeared~\cite{uchino2026erroranalysis}.

%% file: sec-bfsdfs.tex
\section{General BFS Placement for Composed Bilinear Algorithms}
\label{sec:bfsdfs}

The V1-vs-V2 question of \S\ref{sec:critical} and its Toom-Cook-$k$ generalization are instances of a more general one: given two composed bilinear algorithms, an outer one calling an inner one, at which level should the single available processor split (a \emph{BFS step}, in CAPS's~\cite{lipshitz2012,ballard2013strassen} terminology) be placed? The answer is a closed-form threshold when the outer algorithm is size-preserving, as 3M and Toom-Cook-$k$ are, and a different, memory-driven criterion when it is size-reducing, as Strassen is; this is why CAPS's own BFS/DFS choice is not an instance of our Theorem~\ref{thm:generalcritical}, despite the similarity.

Let $\mathcal{A}$ compute one product via $m_A$ auxiliary bilinear products (its \emph{arity}), each formed from fixed linear combinations of $r_A$ \emph{channel} blocks of the original operands, recombined in $O(N^2)$: Toom-Cook-$k$ splits into $r_A=k$ channels and $m_A=2k-1$ products, with Gauss's identity the $k=2$ case ($r_A=2,m_A=3$). Let $\mathcal{B}$ realize each auxiliary product, with bandwidth exponent $\theta_B$ on $P$ processors. \textbf{BFS-$\mathcal{A}$} partitions $P$ into $m_A$ groups of $P/m_A$, each computing one product via $\mathcal{B}$ concurrently (exactly V1 for Toom-Cook-$k$); \textbf{Fuse-into-$\mathcal{B}$} keeps all $P$ processors together, realizing $\mathcal{B}$'s schedule once and moving all $r_A$ channels together (exactly V2).

\begin{theorem}[General BFS placement, size-preserving $\mathcal{A}$]
\label{thm:generalbfs}
For size-preserving $\mathcal{A}$ of arity $m_A$, channel count $r_A$, composed with any $\mathcal{B}$ of communication exponent $\theta_B$, BFS-$\mathcal{A}$ communicates less than Fuse-into-$\mathcal{B}$ iff $\theta_B<\theta_A^\ast:=\log_{m_A}(r_A)$.
\end{theorem}
\begin{proof}
As in Theorem~\ref{thm:invariance}, both strategies perform identical total arithmetic, so the comparison is bandwidth-only. Fuse-into-$\mathcal{B}$ moves $r_A$ channels in one pass (coefficient $r_A$); BFS-$\mathcal{A}$ uses $P/m_A$ processors per group (coefficient $m_A^{\theta_B}$). Solving $m_A^\theta=r_A$ gives $\theta=\log_{m_A}r_A$.
\end{proof}
This recovers Theorem~\ref{thm:generalcritical} ($m_A=2k-1,r_A=k$) and Theorem~\ref{thm:critical} ($k=2$) as special cases, and, instantiated at $m_A=3,r_A=2$, reproduces Table~\ref{tab:classical}'s V1-vs-V2 savings exactly.

Theorem~\ref{thm:generalbfs} assumes $\mathcal{A}$ is size-preserving. Strassen's recursive step is not: it reduces a size-$N$ product to $m_A=7$ size-$(N/2)$ products, so the relevant comparison is BFS ($7$ concurrent groups of $P/7$) versus DFS (the $7$ subproblems sequentially on the full $P$). An analogous calculation shows BFS's bandwidth is $m_A^{\theta_B-1}$ times DFS's, so BFS communicates no more whenever $\theta_B<1$, true for every algorithm here. This explains CAPS's own BFS/DFS choice~\cite{lipshitz2012,ballard2013strassen}: BFS is communication-favorable essentially unconditionally for a size-reducing outer algorithm, so what CAPS trades off is \emph{memory}, using BFS whenever the budget allows $m_A$-fold memory and DFS only when memory-constrained, a criterion orthogonal to Theorem~\ref{thm:generalbfs}'s.

Combining both cases: place BFS at $\mathcal{A}$ if it is size-preserving and $\theta_B<\theta_A^\ast$; otherwise fuse $\mathcal{A}$ and ask the identical question one level deeper, of $\mathcal{B}$ itself. Our outer algorithm (3M or Toom-Cook-$k$) is always size-preserving, composed with an inner one (classical or Strassen) that need not be, so this recursion terminates in one step: the BFS step belongs \emph{before} the matrix multiplication, not inside it, even when the inner algorithm is Strassen, since \S\ref{sec:strassen}'s $\theta_{2D}(\omega_0),\theta_{3D}(\omega_0)$ are exactly the $\theta_B$ that Theorem~\ref{thm:generalbfs} takes as input.

%% file: sec-discussion.tex
\section{Discussion}
\label{sec:discussion}
\label{sec:limitations}
\label{sec:conclusion}

We decompose the question of how to parallelize the 3M complex-matrix-multiplication trick into two independent sub-questions: is 3M worth doing at all (V2 vs.\ V3, purely arithmetic, since both share an identical communication schedule), and is it worth restructuring communication to get more of it (V1 vs.\ V2, purely communication, since both share an identical compute time). For classical matrix multiplication, 3M is unconditionally worth doing locally, by a $25\%$ arithmetic margin matching BLIS's design, and additionally worth splitting into concurrent processor groups for 2D layouts at scale, but never for 3D. This 2D-vs-3D split, however, is governed not by the bandwidth-based critical exponent $\theta^\ast=\log_32$ that first appears to decide it, but by which term, bandwidth or latency, dominates as $P\to\infty$: bandwidth shrinks with $P$ while latency grows, in both layouts, so latency eventually dominates absolutely, regardless of $\omega$. For 2D, this gives V1 a universal $1-1/\sqrt3\approx42.3\%$ latency advantage for \emph{every} arithmetic exponent (versus a merely $13.4\%$ bandwidth advantage for classical specifically, which does depend on $\omega$ and is not universal), not just those with $\theta(\omega)<\theta^\ast$; $\theta^\ast$ instead governs how large $P$ must be before this advantage appears. For 3D, the same latency-growth argument applies, but the fixed cost of V1's redistribution step narrowly outweighs latency's fixed advantage there, so V2 remains asymptotically at least as good, for every $\omega\le3$, the gap vanishing as $P\to\infty$ rather than reversing. The same mechanism generalizes: for Toom-Cook-$k$ (Corollary~\ref{cor:trueasymptotic-genk}), the 2D result generalizes exactly ($1-1/\sqrt{2k-1}$, $\omega$-independent), and the true 3D crossover moves from $k=2$ to $k=3$, also $\omega$-independent; for general replication $c$ (Corollary~\ref{cor:trueasymptotic-genc}), the true asymptotic winner is V1 throughout the entire interior of the 2.5D range, not just up to $c^\ast$, though the $P$ required to observe this grows unboundedly as $c\to P^{1/3}$, so $c^\ast$ remains the practically relevant threshold at any finite scale.

This analysis is entirely model-based, and we have not validated it against a real distributed implementation. The $\alpha$-$\beta$-$\gamma$ model itself omits effects of the same order of magnitude as the percentages we report: contention among V1's three concurrent groups sharing one interconnect; load imbalance; the gap between $\alpha,\beta,\gamma$ as idealized constants and a real collective-communication implementation (scatter, gather, broadcast-reduce); and the $O(n^2)$ lower-order terms dropped throughout, not separately confirmed to remain lower-order at our crossover scales. $T_{\text{redist}}$ in particular is a coarse estimate, not derived from a concrete collective schedule. An MPI-based implementation on an existing 2D/2.5D/3D library, or on CAPS directly, would calibrate these terms and validate the crossover points~\eqref{eq:general-condition} against measurement; we leave this to future work. Short of that, our consistency-check simulator (supplementary material, \S\ref{sec:results}) verifies the closed forms are algebraically consistent with an independent, discrete accounting of the same model, which is necessary but not sufficient for the model to be accurate, and does not address the omissions above.

The two kinds of claim this paper makes carry different dependence on $\gamma,\beta,\alpha$. The \emph{structural} claims, including the sign of every asymptotic crossover (V1 beats V2 in 2D, loses in 3D, for every arithmetic exponent; $k=3$ rescues 3D for Toom-Cook-$k$), are $P\to\infty$ limits of latency-term ratios alone (Lemma~\ref{lem:latencydominance}), and $\gamma,\beta,\alpha$ cancel out entirely, not appearing in the stated ratios ($1/\sqrt3$, $\log_23$ vs.\ $2$) at all; these are robust to the choice of machine constants. What does depend on $\gamma,\beta,\alpha$, via $T_{\text{redist}}$ and the bandwidth term, is \emph{where} the crossover occurs: the finite-$P$ threshold $c^\ast$ (Corollary~\ref{cor:criticalc}) and the exact percentages in Table~\ref{tab:numericcomparison} and \S\ref{sec:results}, computed at one representative setting and shifting under another. Treat the asymptotic direction of each crossover as the robust contribution, the percentages as illustrations at one operating point.

Proposition~\ref{prop:memory}'s $6$-vs-$8$-block memory comparison is a distributed-memory accounting and does not transfer to BLIS's single-node implementation~\cite{vanzee2017complex}: BLIS packs operands into small, cache-blocked workspace buffers rather than holding each as one persistent block, so whatever extra workspace its 3m method needs is transient and cache-block-sized, not a full extra block.

For Strassen's algorithm, using the Karstadt-Schwartz addition count, this picture inverts: in-place 3M is strictly worse than standard, since its recombination phase cannot be shared across channels the way evaluation can, making standard 4M best for 3D-Strassen at every scale we evaluated. Two assumptions bear on this: the latency exponent for Strassen- and CAPS-based layouts is assumed identical to classical, and 3M's known numerical-stability disadvantage~\cite{stability3m} is not in the cost model, which may affect V1/V2's viability for ill-conditioned inputs; error bounds for alternative-basis Strassen~\cite{schwartz2024altbasis} could sharpen this. Whether other fast algorithms (Pan-, Schönhage-type) show the same V2-vs-V3 reversal is also worth checking.

The framework generalizes beyond Gauss's identity: applied to Karatsuba's high/low-digit split for FP64-via-low-precision emulation, the identical two theorems govern the same comparison, correctly predicting a classical-like ranking, since that identity has no deep recursive recombination. The same stability caveat applies here too, more acutely: this recombination has not, to our knowledge, been separately checked for stability.

Generalizing to Toom-Cook-$k$ yields one closed-form superiority condition, combining the arithmetic saving $(k-1)^2/k^2$ with the critical exponent $\theta^\ast(k)=\log_{2k-1}k$, both increasing in $k$: where $k=2$ leaves 3D on the wrong side, $k=3$ flips it favorable for \emph{every} arithmetic exponent asymptotically, though for Strassen, where $\theta^\ast(3)$ still falls short of $\theta_{3D}(\omega_0)$, this only appears at larger $P$ than the bandwidth-only threshold suggests (\S\ref{sec:critical}). A further direction, suggested by Nissim et al.'s alternative-basis and Toom-Graph techniques~\cite{nissim2024toomcook}, is optimizing the $O(n^2)$ lower-order terms suppressed throughout, instead of treating them as negligible.

A separate open direction is fault tolerance: how can a fault-tolerant complex matrix multiplication be built on V1, V2, or V3? Fault-tolerant matrix multiplication was already established~\cite{huangabraham1984,bosilca2009abft,chen2008failstop,birnbaum2018ftmm,birnbaum2020ftmm}, and fault-tolerant recursive Toom-Cook likewise~\cite{chen2024faulttoomcook,park2025faulttoomcook}, but neither addresses this composition. Adapting either would mean deciding whether the fault-tolerance layer sits on the outer 3M split, the inner algorithm, or both, and whether V1's already-disjoint groups can double as the needed redundancy structure.